\documentclass[10pt,conference]{IEEEtran}
\IEEEoverridecommandlockouts
\usepackage{cite}
\usepackage{amsmath,amssymb,amsfonts}
\usepackage{algorithmic}
\usepackage{graphicx}
\usepackage{textcomp}
\usepackage[hidelinks]{hyperref}
\usepackage{float} 
\usepackage{xcolor}
\usepackage[most]{tcolorbox}
\usepackage{array}
\newcolumntype{C}[1]{>{\centering\arraybackslash}p{#1}}
\def\BibTeX{{\rm B\kern-.05em{\sc i\kern-.025em b}\kern-.08em
    T\kern-.1667em\lower.7ex\hbox{E}\kern-.125emX}}

\usepackage[pscoord]{eso-pic}

\AddToShipoutPicture* {%
  \AtPageLowerLeft{%
    \hspace{0.35\paperwidth}% Adjust left position
    \raisebox{0.03\paperheight}{% Adjust bottom height
      \fontsize{8}{10}\selectfont
      \noindent Accepted for publication at IEEE Quantum Week 2026.
    }%
  }%
}
\begin{document}

\title{Adaptive Error Budget Allocation for Fault-Tolerant Quantum Resource Estimation: A Metaheuristic Approach}

\author{\IEEEauthorblockN{Asif Akhtab Ronggon}
\IEEEauthorblockA{\textit{Division of Computer Science and Engineering} \\
\textit{Louisiana State University}\\
Baton Rouge, USA \\
arongg1@lsu.edu}
\and
\IEEEauthorblockN{Tasnuva Farheen}
\IEEEauthorblockA{\textit{Division of Computer Science and Engineering} \\
\textit{Louisiana State University}\\
Baton Rouge, USA \\
tfarheen@lsu.edu}

}

\maketitle

\begin{abstract}
System-level resource estimation is a key component of fault-tolerant quantum computing~(FTQC) toolchains. Its efficiency depends on how global error tolerance is allocated across logical operations, T-state distillation, and rotation synthesis to minimize physical resource overhead. The commonly used uniform-allocation strategy ignores circuit-specific structure and can overprovision inactive or less critical subsystems, leading to inflated space-time estimates. Prior work aims to address this limitation using supervised models trained on offline-generated datasets. However, this approach incurs additional data-generation costs and limits deployment flexibility. To overcome these drawbacks, we propose a training-free optimization framework that performs derivative-free search directly on the Azure Quantum Resource Estimator~(AQRE), enabling instance-specific error budget allocation for previously unseen circuits without requiring offline training data. To evaluate robustness to optimizer choice, we instantiate the framework with two structurally distinct metaheuristics, simulated annealing and quantum particle swarm optimization. We evaluate our framework across 433 circuits spanning 2 to 91 qubits from 31 families in the MQT Bench suite. Across the benchmark suite, both methods reduce space-time cost by more than 33\% on average and agree within 1.34\% points, indicating that the gains are stable across different metaheuristic search strategies. Our analysis further finds that the optimization benefit is driven primarily by error-profile asymmetry rather than circuit scale, and the metric, Gini coefficient of optimized allocation, provides an interpretable diagnostic of expected improvement. Together, these results position adaptive error budget allocation as a system-software optimization layer for FTQC resource estimation pipeline.

\end{abstract}


\section{Introduction}

Quantum computing has advanced rapidly in both hardware and software stacks, driving a growing interest in applications ranging from healthcare \cite{flother2023state,bhavin2021blockchain} and  chemistry\cite{yamamoto2025quantum,motta2022emerging,lee2023evaluating} to combinatorial optimization \cite{kim2025quantum,ponce2025graph} and finance \cite{herman2023quantum}. To implement these applications, computational problems must be encoded as quantum circuits that produce the desired solutions when executed. However, quantum hardware remains highly error-prone, with error rates that increase as circuit depth and width grow. This limits the reliability of large-scale quantum computations. Fault-tolerant quantum computing addresses this fundamental barrier through quantum error correction\cite{kecceci2025quantum}. However, it imposes a substantial resource overhead that transforms the compilation process into a critical bottleneck in the quantum design automation workflow, governing whether algorithmic designs can be mapped onto physically realizable hardware within acceptable space–time bounds. 

As quantum computing advances beyond the noisy intermediate-scale quantum (NISQ) era toward circuits comprising millions of logical gates, minimizing the space-time volume, defined as the product of the number of physical qubits and the algorithmic runtime, becomes a critical requirement for scalable implementation. A key challenge in this context is to allocate the global error budget across three functionally distinct and resource-dependent subsystems: logical operations, T-state distillation, and rotation synthesis via gate decomposition~\cite{van2023using,microsoft2026resourceestimator}. Each subsystem exhibits fundamentally different scaling characteristics with respect to its assigned error budget. The prevalent strategy of uniform error budget allocation fails to exploit the structural characteristics of individual circuits, often resulting in over-provisioning of error-insensitive subsystems and under-provisioning of critical bottlenecks. This misallocation leads to unnecessary inflation of the overall resource requirements for fault-tolerant quantum computation.

Recent advances have sought to overcome the limitations of uniform error allocation by introducing strategies for distributing error budgets. Forster et al.\cite{forster2025improving} proposed a supervised learning framework that predicts allocation vectors based on AQRE-derived circuit features, achieving reductions of up to 77.7\% for individual circuits and 15.6\% on average across 383 MQT Bench instances\cite{quetschlich2023mqtbench}. Asif et al.\cite{ronggon2026gametheoreticapproachoptimizing} formulated error budget allocation as a game-theoretic (GT) optimization problem and reported a 30.22\% average reduction across a 433-circuit MQT Bench. These results indicate that non-uniform allocation can substantially reduce space-time overhead. However, both approaches present notable limitations. The supervised learning method requires substantial offline data generation and training costs. The GT approach depends on a specific equilibrium-based search dynamic, and the resulting solution may not correspond to the optimal allocation under the estimator objective. Therefore, there remains a need for a practical estimator-side optimization technique that can provide substantial improvements across a broad range of circuits and integrate efficiently with fault-tolerant resource estimation and compilation workflows.

We address these limitations by applying \textit{derivative-free metaheuristic optimization techniques} directly to the resource estimation process. This approach removes the reliance on offline training data, generalizes to arbitrary circuits without prior exposure, and does not constrain the optimization to a single equilibrium-based search strategy. We instantiate this methodology with two structurally distinct optimizers: simulated annealing (SA) and quantum particle swarm optimization (QPSO). We evaluated both approaches on 433 circuit instances spanning 31 circuit families from the MQT Bench. Our study is structured around the following research questions:

\noindent\hypertarget{rq1}{\underline{\textbf{RQ1.}}}  Can an optimization method that requires neither offline training nor gradient information consistently improve fault-tolerant quantum resource estimation over the uniform error budget baseline in a diverse benchmark suite?

\noindent\hypertarget{rq2}{\underline{\textbf{RQ2.}}} Which factors explain the variations in improvement among quantum circuit families under optimized error budget allocation?

\noindent\hypertarget{rq3}{\underline{\textbf{RQ3.}}} To what extent do structural circuit features such as the number of logical qubits, the T-count, and the rotation count explain or predict the observed improvement?

\noindent\hypertarget{rq4}{\underline{\textbf{RQ4.}}} How consistent are the resulting allocations and resource improvements across different metaheuristic optimizers?

\noindent \textbf{Contributions.} Our main contributions are summarized below.

\begin{itemize}

\item We develop a software-layer optimization framework that handles error budget allocation as an instance-specific optimization step within the fault-tolerant resource estimation pipeline, without requiring offline training data or gradient information.

\item We show that the proposed framework significantly reduces the estimated space-time overhead across 433 circuits from 31 benchmark families, with mean improvements of $35.18\pm0.11\%$ using Simulated Annealing and $33.84\pm0.08\%$ for Quantum Particle Swarm Optimization.

% averaged over four random seeds.

\item We demonstrate that these gains remain consistent across the two tested metaheuristic optimizers, with average improvements differing by only 1.34 percentage points, suggesting that the observed benefit is not tied to a single search strategy.

\item We analyze improvement variation across circuit families and find that error-profile asymmetry better explains optimization benefits than structural features.

\end{itemize}

The remainder of this paper is organized as follows.
Section~\ref{preliminaries} reviews quantum error correction and resource estimation. Section~\ref{problem} formalizes the allocation problem. Sections~\ref{sa} and~\ref{qpso} present the SA and QPSO frameworks respectively. Section~\ref{experimental} describes the experimental setup. Section~\ref{result} reports the empirical evaluation. Section~\ref{conclusion} concludes.

\section{Preliminaries}
\label{preliminaries}

\subsection{Quantum Error Correction}\label{AA}
Fault-tolerant quantum computation requires encoding logical qubits into a larger number of physical qubits using a quantum error-correcting code, such that errors can be detected and corrected without destroying the encoded quantum information\cite{gottesman1998theory,katabarwa2024early}. The surface code\cite{fowler2012surface,bravyi1998quantum} has emerged as the leading candidate architecture for fault-tolerant quantum computation, primarily because it requires only nearest-neighbor interactions on a two-dimensional qubit lattice while maintaining a comparatively high error-correction threshold of approximately 1\% per physical gate operation \cite{stephens2014fault}. This combination of architectural simplicity and noise tolerance makes it particularly well-suited for practical implementations of quantum error correction and, consequently, for the error budget optimization framework developed in this work.
\subsection{Resource Estimation}\label{BB}
To execute a fault-tolerant quantum algorithm within an allowed overall error budget $\varepsilon$, the error budget must account for all sources of error throughout the computation. Following the framework established by Beverland et al.~\cite{beverland2022assessing} and implemented in the AQRE~\cite{van2023using}, the total error budget is partitioned into three components:
\begin{equation}
\varepsilon_{\text{total}} = \varepsilon_{\text{logical}} + \varepsilon_{\text{tstates}} + \varepsilon_{\text{rotations}}
\end{equation}

where $\varepsilon_{\text{logical}}$ denotes the portion of the error budget reserved for logical qubit operations; $\varepsilon_{\text{tstates}}$ represents the portion allocated to producing reliable magic states through distillation; and $\varepsilon_{\text{rotations}}$ corresponds to the portion assigned to approximating arbitrary rotations through Clifford+T decompositions\cite{van2023using,microsoft2026resourceestimator}. These allocations are interdependent within the resource estimation pipeline, requiring careful balancing to minimize hardware overhead while maintaining the desired computational reliability\cite{forster2025improving}.

\section{Problem Formulation}
\label{problem}
\subsection{Error Budget Allocation as Constrained Optimization}

The allocation of the error budget is based on a constrained optimization problem on the probability simplex. Given a total error budget $\varepsilon_{\text{total}}$ representing the maximum tolerable failure probability for the quantum algorithm, we seek an allocation vector $\mathbf{s} = (s_L, s_T, s_R)$ that distributes error fractions among three functionally distinct subsystems: logical operations ($L$), distillation of the T-state ($T$) and rotation synthesis ($R$). The allocation is constrained to the $\varepsilon$-interior of the 2-simplex:

\begin{equation}
\label{simplex}
    \mathbf{S}_{\varepsilon} = \left\{ \mathbf{s} \in \mathbb{R}^3_+ \;\middle|\; \sum_{i} s_i = 1,\; s_i \geq \varepsilon \right\}
\end{equation}

where $\varepsilon$ ensures that no subsystem is assigned a degenerate allocation that would cause the resource estimator to produce infeasible configurations. Under this constraint, $\mathbf{S}_{\varepsilon}$ forms a truncated version of the standard 2-simplex, resulting in a compact two-dimensional feasible region over which optimization is performed, with each subsystem receiving a corresponding fraction of the total error. The AQRE then acts as a black-box mapping $\mathbf{s} \mapsto (Q(\mathbf{s}), R(\mathbf{s}))$, where the resulting resource estimates vary in a piecewise manner due to the underlying discrete design choices.

\subsection{Cost Landscape Characterization}
\label{landscape}
The optimization problem formulated by the resource estimation model is inherently non-convex, discontinuous, and structurally disordered as the resource estimator takes continuous allocation input to physical resources through discrete internal decisions, such as integer-valued code distance choice and finite T-factory design catalogs. Minimal changes in the error allocation vector can induce sudden change of the regime, causing step-wise variations in the number of physical qubits and execution time. As a result, the objective function exhibits a rugged, multimodal surface with a large number of local minima and no cost separations. These properties have a negative impact on the use of gradient-based optimization algorithms, which assume smoothness and differentiability at the edges of regimes that is not met, and restrict the utility of deterministic local search algorithms that do not have escape mechanisms of suboptimal basins of attraction. Additionally, the simplex constraint of allocation vector also restricts possible search space and thus any optimization algorithm must ensure that its search remains budget feasible at each step.

% Simulated annealing (SA)\cite{van1987simulated,busetti2003simulated} fits this environment perfectly, working directly on black-box, non-smooth objective functions and including a probabilistic acceptance criterion allowing controlled movement over cost barriers between local minima. Its exploration strategy based on temperature toleration allows extensive searching all over the world during the initial high temperature stage and increases concentration around high-quality areas as the temperature drops, thus offering an efficient way to navigate the extremely rugged cost terrain. SA, therefore, provides a conceptually well-founded and methodologically sound way of optimizing error budget balancing under the conditions of discrete and non-convex resource estimation dynamics\cite{rajwar2023exhaustive}.

\subsection{Optimization Objective}

The resource estimator returns two competing quantities for any allocation: physical qubit count $Q(\mathbf{s})$ and algorithmic runtime $R(\mathbf{s})$. Optimizing either one in isolation is insufficient, as reducing qubits often increases runtime and vice versa. We therefore define a composite cost function that jointly captures space--time complexity:
\begin{equation}
\mathcal{J}(\mathbf{s}) = Q(\mathbf{s})^{\alpha} \cdot R(\mathbf{s})^{1 - \alpha}
\label{eq:cost}
\end{equation}
where $\alpha \in [0,1]$ controls the trade-off between qubit count and runtime. This formulation corresponds to a geometric mean objective, ensuring scale-invariant balancing between competing resource dimensions. In practice, we use $\alpha = 0.5$, yielding:
\begin{equation}
\mathcal{J}(\mathbf{s}) = \sqrt{Q(\mathbf{s}) \cdot R(\mathbf{s})}
\label{eq:cost_half}
\end{equation}
which is monotonically equivalent to minimizing the space--time volume $Q \cdot R$, the standard metric in fault-tolerant resource estimation. The optimization problem is therefore:
\begin{equation}
\mathbf{s}^* = \arg\min_{\mathbf{s} \in \mathcal{S}_{\varepsilon}} \mathcal{J}(\mathbf{s})
\label{eq:opt}
\end{equation}
where \(\mathbf{s}^*\) denotes the allocation that achieves the minimum value of the estimator-induced space-time objective within the feasible simplex \(\mathcal{S}_{\varepsilon}\).

\section{SIMULATED ANNEALING FRAMEWORK}
\label{sa}
\begin{figure*}[!t]
    \centering
    \includegraphics[width=1\linewidth]{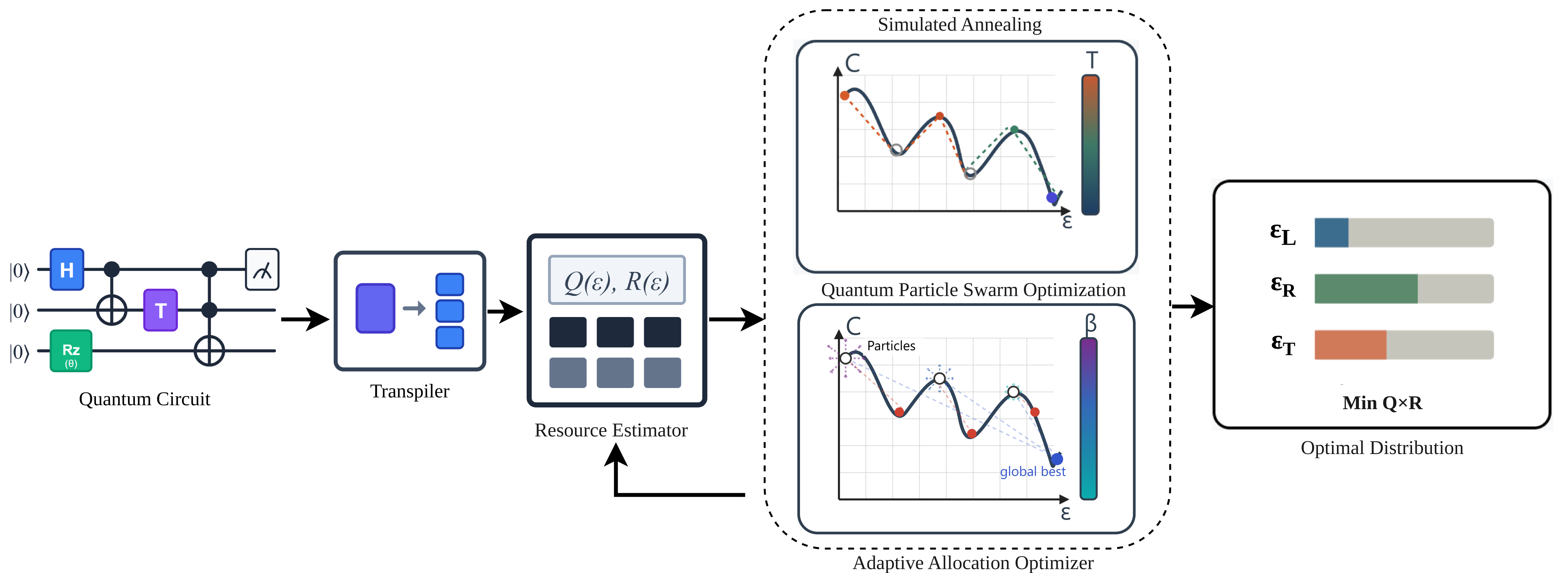}
    \caption{End-to-end pipeline: quantum circuits are transpiled, characterized through the AQRE, and optimized through SA or QPSO to minimize the space-time product $Q \times R$.}
    \label{fig:methodology}
\end{figure*}

Simulated annealing (SA) offers a principled solution to the error budget allocation problem by enabling direct optimization of black-box, non-smooth objective functions. The probabilistic acceptance mechanism inherent to SA allows the algorithm to overcome cost barriers associated with local minima~\cite{van1987simulated,busetti2003simulated}. To systematically address the allocation problem, we develop an SA-based framework that explores the allocation simplex using four key components: an adaptive perturbation strategy to efficiently navigate the search space, a cost-normalized acceptance criterion to ensure appropriate acceptance probabilities, an adaptive cooling schedule to control convergence, and a multi-start initialization scheme to enhance coverage of the solution space. The temperature parameter modulates the extent of exploration, with higher values promoting broader search and lower values concentrating the search in regions of interest. As a result, SA provides an effective approach for optimizing error budget allocation within the discrete and non-convex landscape characteristic of fault-tolerant resource estimation~\cite{rajwar2023exhaustive}. Figure~\ref{fig:methodology} illustrates the end-to-end pipeline.

\subsection{State Transition}

At each iteration, the algorithm generates a candidate allocation by perturbing the current solution. The perturbation must be large enough to escape local plateaus in the early stages of the search, yet small enough to refine promising allocations in the later stages. We achieve this by tying the step size to both the current temperature and the fraction of iterations completed.

At iteration $k$, a neighbor $\mathbf{s}'$ is generated from the current allocation $\mathbf{s}^{(k)}$ by:
\begin{equation}
\mathbf{s}' = \text{Proj}_{\mathcal{S}_{\varepsilon}} \left( \mathbf{s}^{(k)} + \boldsymbol{\delta}^{(k)} \right)
\label{eq:neighbor}
\end{equation}
where each perturbation component is drawn independently as:
\begin{equation}
\delta_i^{(k)} \sim \mathcal{U}(-\eta_k,\; \eta_k)
\label{eq:perturbation}
\end{equation}

The step size $\eta_k$ starts large and shrinks as the search progresses, ensuring broad exploration early and focused refinement later. The projection operator $\text{Proj}_{\mathcal{S}_{\varepsilon}}$ maps the perturbed vector back onto the feasible simplex $\mathcal{S}_{\varepsilon}$ by clamping each component to a minimum of $\varepsilon$ and renormalizing to unit sum. This guarantees that every candidate allocation is feasible without requiring any auxiliary constraint-handling mechanism.

\subsection{Acceptance Criterion}

A deterministic acceptance of all candidate solutions is undesirable. Although cost reducing moves are always beneficial, the algorithm must occasionally permit higher cost transitions to escape local minima and ensure adequate exploration of the solution space. The standard Metropolis criterion $\exp(-\Delta/T)$ \cite{metropolis1953equation} controls this trade-off through temperature $T$. However, this standard form assumes that cost magnitudes are comparable across problem instances. In our setting, a small circuit may cost $10^6$, while a large circuit reaches $10^{15}$, making a fixed temperature meaningless across the benchmark suite.

We resolve this by normalizing the cost difference by the current cost value. Let $\Delta = \mathcal{J}(\mathbf{s}') - \mathcal{J}(\mathbf{s}^{(k)})$ denote the cost difference. The acceptance rule is:
\begin{equation}
P(\text{accept}) = 
\begin{cases} 
1, & \text{if } \Delta < 0 \\[6pt]
\exp\left(-\dfrac{\Delta}{T_k \cdot \mathcal{J}(\mathbf{s}^{(k)})}\right), & \text{otherwise}
\end{cases}
\label{eq:accept}
\end{equation}

The denominator $T_k \cdot \mathcal{J}(\mathbf{s}^{(k)})$ means the algorithm evaluates the relative cost increase $\Delta / \mathcal{J}(\mathbf{s}^{(k)})$ rather than the absolute value of $\Delta$. A 5\% cost increase is treated the same way whether the circuit costs $10^6$ or $10^{15}$, producing consistent exploration behavior across the entire benchmark suite.

\subsection{Adaptive Cooling Schedule}

The temperature $T$ controls how willing the algorithm is to accept worse solutions. High temperatures encourage exploration; low temperatures enforce exploitation. The rate at which $T$ decreases therefore determines how quickly the algorithm transitions between these two regimes.

A fixed cooling rate treats all phases identically, regardless of whether the algorithm is actively finding improvements or stagnating. Adaptive cooling schedules that adjust based on search progress have been proposed to improve efficiency \cite{karabin2020simulated}. We instead adapt the rate based on the recent improvement rate $\rho_k$, defined as the relative cost decrease of the most recent accepted improving move. The temperature update is:
\begin{equation}
T_{k+1} = T_k \cdot \gamma(\rho_k)
\label{eq:cool}
\end{equation}
where the cooling factor $\gamma$ responds to the optimization dynamics:
\begin{equation}
\gamma(\rho_k) = 
\begin{cases} 
0.99, & \rho_k > \tau_{\text{high}} \\
0.96, & \rho_k < \tau_{\text{low}} \\
0.98, & \text{otherwise}
\end{cases}
\label{eq:gamma}
\end{equation}

When the algorithm is finding large improvements ($\rho_k > \tau_{\text{high}}$), cooling slows down to give the search more time in the productive region. When improvement has stalled ($\rho_k < \tau_{\text{low}}$), cooling speeds up to push the algorithm toward convergence rather than wasting iterations. To prevent the temperature from dropping too low too early, periodic reheating is applied:
\begin{equation}
T_k \leftarrow 1.05 \cdot T_k \quad \text{every } K \text{ iterations}
\label{eq:reheat}
\end{equation}

This brief temperature increase re-enables uphill moves at regular intervals, allowing the algorithm to escape plateaus it may have committed to prematurely.

\subsection{Multi-Start Initialization via Latin Hypercube Sampling}

A single SA run will converge to whatever local minimum lies in the basin of its starting point. If the starting point happens to be near a poor local minimum, the final solution will be poor regardless of how well the cooling schedule is tuned. To reduce this dependence on initialization, we run SA multiple times from different starting points and keep the best result.

The starting points are generated using Latin Hypercube Sampling (LHS)\cite{mckay2000comparison}, which partitions each dimension of the unit hypercube into equal-probability intervals and places exactly one sample per interval:
\begin{equation}
\mathbf{s}_j^{(0)} \sim \text{LHS}(n,\; d=3)
\label{eq:lhs}
\end{equation}

The samples are then mapped onto the probability simplex through exponential normalization:
\begin{equation}
\tilde{s}_i = \frac{e^{x_i}}{\sum_j e^{x_j}}
\label{eq:softmax}
\end{equation}

This produces starting points that are spread uniformly across the allocation space, unlike Dirichlet sampling which concentrates points near the simplex center. The globally best allocation found across all $n$ restarts is returned as the final optimized solution.

\subsection{Convergence Criterion}

Every restart does not need to run for the full iteration budget. Once the search has stabilized and the cost is no longer changing, continuing to iterate wastes computation that could be used by other restarts. We detect this stabilization by monitoring the recent history of accepted cost values. Let $W$ denote a sliding window over the most recent accepted costs. Convergence is declared when the relative variation within this window falls below a threshold $\delta$:
\begin{equation}
\frac{\max_{k \in W} \mathcal{J}_k - \min_{k \in W} \mathcal{J}_k}{\max_{k \in W} \mathcal{J}_k} < \delta
\label{eq:converge}
\end{equation}

When this condition is met, the accepted costs are effectively constant, indicating that the search has settled near a local minimum. The current restart terminates early, and the saved iterations are available for subsequent restarts that may still be exploring productive regions.

\subsection{Baseline Comparison}

To quantify the benefit of optimized allocation, all results are compared against the uniform distribution baseline, which is the default setting of the AQRE:
\begin{equation}
\mathbf{s}_{\text{uniform}} = \left(\frac{1}{3},\; \frac{1}{3},\; \frac{1}{3}\right)
\label{eq:uniform}
\end{equation}

Performance improvement is measured as the relative reduction in the space-time cost:
\begin{equation}
\text{Improvement} = \frac{\mathcal{J}(\mathbf{s}_{\text{uniform}}) - \mathcal{J}(\mathbf{s}^*)}{\mathcal{J}(\mathbf{s}_{\text{uniform}})} \times 100\%
\label{eq:improvement}
\end{equation}

A positive value indicates that SA optimized allocation requires fewer physical resources than the uniform baseline, with larger values corresponding to greater resource savings.

\section{Quantum Particle Swarm Optimization}
 \label{qpso}
As an alternative search paradigm, we apply Quantum Particle Swarm Optimization (QPSO) to the same allocation problem defined in Section \ref{problem}. While SA escapes local minima through thermal excitation, where the Metropolis criterion in Eq. \eqref{eq:accept} accepts uphill moves with probability proportional to the cost difference, QPSO employs a fundamentally different mechanism rooted in quantum tunneling. In QPSO, each particle is modeled as occupying a quantum state within a Delta potential well, and its position is sampled from the resulting probability distribution rather than updated through deterministic velocity equations\cite{sun2004global,sun2006quantum}. We apply QPSO to the cost function $\mathcal{J}(\mathbf{s})$ defined in Eq. \eqref{eq:cost} over the feasible region $\mathcal{S}_\varepsilon$ defined in Eq.\eqref{simplex}, and compare all results against the uniform baseline using the improvement metric in Eq. \eqref{eq:improvement}.

\subsection{Position Update}

A swarm of $N$ particles searches over generations $k = 1, \dots, K$.
Let $\mathbf{x}_i^{(k)} \in \mathbb{R}^D$ denote the position of
particle $i$ at generation $k$, $\mathbf{p}_i$ its personal best
position, and $\mathbf{g}$ the global best across the swarm. For each
coordinate~$d$, a stochastic attractor is formed as the random convex
combination

\begin{equation}
  c_{i,d} = \varphi_{i,d} \, p_{i,d}
           + (1 - \varphi_{i,d}) \, g_d,
  \quad \varphi_{i,d} \sim \mathcal{U}(0,1)
  \label{eq:attractor}
\end{equation}

The position update, derived from the inverse cumulative distribution of a particle bound in a Delta potential well~\cite{sun2004global}, is

\begin{equation}
  x_{i,d}^{(k+1)} = c_{i,d}
    \;\pm\; \beta_k \, \bigl\lvert \bar{p}_d - x_{i,d}^{(k)} \bigr\rvert
    \, \ln\!\frac{1}{u_{i,d}},
  \quad u_{i,d} \sim \mathcal{U}(0,1)
  \label{eq:qpso_update}
\end{equation}

where the sign is chosen with equal probability and
$\bar{p}_d = N^{-1}\sum_{i=1}^{N} p_{i,d}$ is the mean best position.
The $\ln(1/u)$ term samples a displacement from a heavy-tailed
distribution whose spatial reach is scaled by the distance from the
particle to the mean best. This allows a particle to traverse an entire plateau and cross the adjacent code-distance boundary in a single step, without requiring gradient information from the piecewise-constant surface described in Section \ref{landscape}.

The contraction--expansion coefficient $\beta_k$ is decreased linearly from $\beta_{\max} = 1.0$ to $\beta_{\min} = 0.5$ across the $K$ generations and plays a role analogous to the temperature in SA. Large values permit long-range tunneling during early exploration, while small values confine particles near their attractors during late refinement. Unlike the adaptive cooling in Eq.~\eqref{eq:cool}, this schedule is deterministic and requires no threshold parameters.

\subsection{Simplex Handling via Log-Ratio Transform}

SA handles the simplex constraint through projection, where the
perturbed vector in Eq.~\eqref{eq:neighbor} is clamped to a minimum of $\varepsilon$ and renormalized. This introduces boundary bias because perturbations near the simplex edges are asymmetrically truncated. We instead map the open 2-simplex onto unconstrained $\mathbb{R}^2$ using the additive log-ratio (ALR) transform\cite{aitchison1982statistical}, taking $s_L$ as
the reference component.

\begin{equation}
  y_1 = \ln\!\frac{s_T}{s_L}, \quad
  y_2 = \ln\!\frac{s_R}{s_L}
  \label{eq:forward}
\end{equation}

The inverse recovers the allocation via softmax over
$\mathbf{z} = (0,\, y_1,\, y_2)$.

\begin{equation}
  s_k = \frac{\exp(z_k)}{\textstyle\sum_j \exp(z_j)}
  \label{eq:inverse}
\end{equation}

This mapping is a diffeomorphism between the open simplex and
$\mathbb{R}^2$. The QPSO updates in
Eqs.~\eqref{eq:attractor}--\eqref{eq:qpso_update} operate entirely in
log-ratio space, and the simplex constraint is satisfied exactly by
construction at every evaluation. No projection, clamping, or
renormalization is required at any step of the algorithm.

After the inverse transform, a minimum fraction guard ensures each
component receives at least $\varepsilon_i = 0.05$ of the total budget. The guard reserves $\varepsilon_i$ per component and distributes the
remaining $1 - 3\varepsilon_i$ proportionally to the transformed fractions. This is not an algorithmic constraint, but a physical one imposed by the resource estimator's code-distance ceiling of $d \leq 50$. It is applied as a post-processing step and does not affect the unconstrained geometry in which QPSO operates.

\subsection{Multi-Restart and Early Stopping}

Each of $n$ independent restarts initializes particle positions by
drawing from the symmetric Dirichlet distribution
$\mathrm{Dir}(1,1,1)$, which is uniform on the simplex, and mapping the samples to the log-ratio space via Eq.~\eqref{eq:forward}. Within each restart, convergence is declared when the global best cost remains unchanged for 30 consecutive generations, after which the restart terminates and the saved evaluations are available to subsequent restarts. The best global allocation in all $n$ restarts is returned as the final solution.

\section{Experimental Setup}
\label{experimental}
The benchmark suite consists of 433 circuit instances from 31 families in MQT Bench~\cite{quetschlich2023mqtbench}, spanning 2 to 91 qubits. Each circuit is transpiled with Qiskit at optimization level~3 and evaluated using AQRE under its default surface code configuration with a physical error rate of $10^{-3}$. All results are measured against uniform baseline allocation $(1/3,\;1/3,\;1/3)$. To assess reproducibility, both optimizers are evaluated across four independent random seeds (42, 90, 60, 210).

The SA optimizer performs at most 200 iterations per restart over $n = 5$ restarts initialized via Latin Hypercube Sampling. The initial temperature is $T_0 = 10.0$ with adaptive cooling factors of 0.99, 0.98 and 0.96. The adaptive cooling thresholds are $\tau_{\text{high}} = 0.01$ and $\tau_{\text{low}} = 0.001$. Restarting occurs early when the relative cost variation within a window of 50 accepted values falls below $5 \times 10^{-4}$. The minimum allocation per component is 0.01. The QPSO optimizer evolves a swarm of 30 particles for at most 150 generations over $n = 5$ restarts initialized via Dirichlet sampling. The contraction expansion coefficient $\beta$ decreases linearly from 1.0 to 0.5, and restarting terminates early after 30 generations without improvement. The minimum allocation per component is 0.05. All experiments run on Kaggle with Python 3.10, Qiskit 2.3.1, MQT Bench, and the .NET SDK 8.0 runtime with the \texttt{qsharp} and
\texttt{azure-quantum} packages for resource estimation.

\section{Results and Analysis}
\label{result}
This section presents the empirical evaluation of the proposed SA and QPSO-based error budget optimizer across the full benchmark suite. The analysis proceeds in three stages: aggregate performance across circuit families, the relationship between structural circuit features and optimization gain, and the role of allocation inequality as a predictive indicator of improvement magnitude.

\subsection{Aggregate Performance Across Circuit Families}

\begin{figure*}[!t]
    \centering
    \includegraphics[width=1\linewidth]{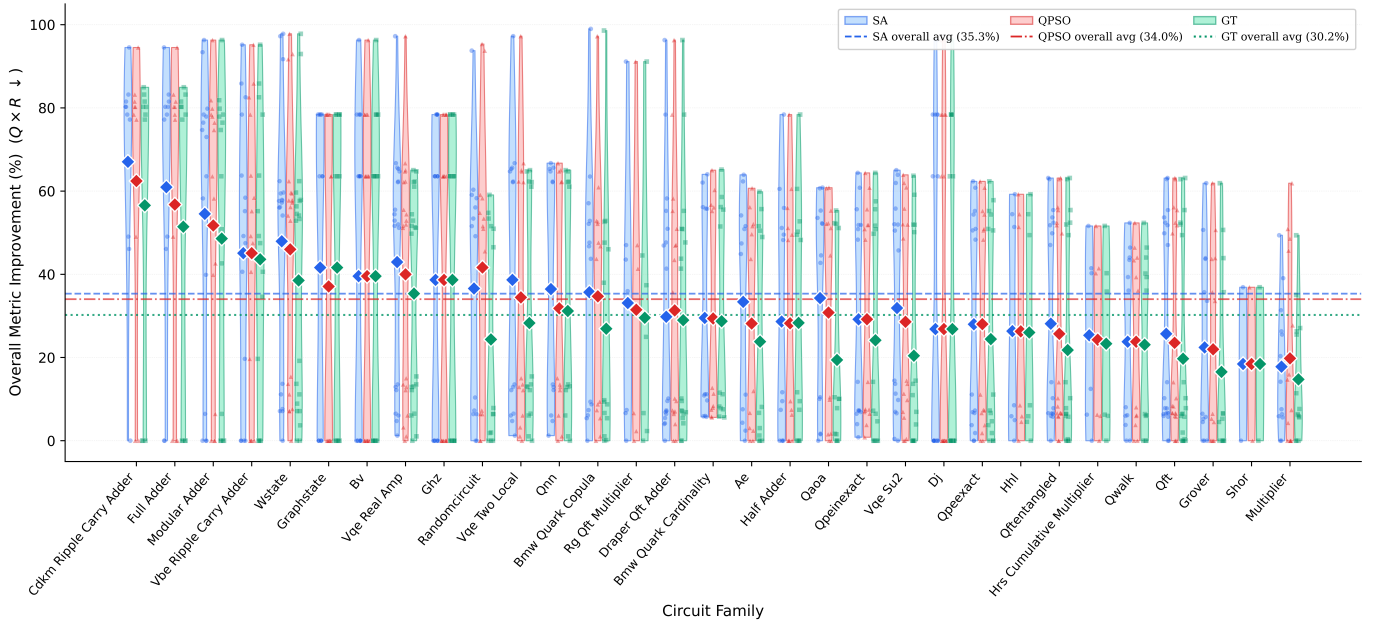}
    % \vspace{-0.1in}
    \caption{Distribution of space-time improvement across 31 circuit families. Dashed lines indicate overall averages for a single representative seed; multi-seed averages are reported in Table \ref{tab:comparison}}
% \vspace{-0.15in}
    \label{fig:violin}
\end{figure*}

Figure~\ref{fig:violin} compares the distribution of space-time improvement $(\mathcal{J}(\mathbf{s}_{\text{uniform}}) -
\mathcal{J}(\mathbf{s}^*))/\mathcal{J}(\mathbf{s}_{\text{uniform}})$ across 31 circuit families for SA, QPSO, and GT method. All per-family
distributions and scatter plots in Figures~\ref{fig:violin}--\ref{fig:gini} are computed from a single representative seed (seed 60) for SA and QPSO; multi-seed statistics are reported in Table~\ref{tab:comparison}. The observed family-level trends are consistent across all three approaches, with the same circuit families consistently ranked at the top and bottom. This consistency suggests that the primary source of variation arises from the circuit-dependent error profile, rather than from optimizer-specific effects. Furthermore, both SA and QPSO increase the family-level averages relative to GT across most benchmarks, indicating that direct metaheuristic search can exploit additional opportunities for improvement while preserving the same underlying structural pattern.

\paragraph{Clifford and T-gate dominated circuits}
The largest improvements are observed in circuit families, where the fault-tolerant cost is dominated by a single mechanism. This pattern is particularly pronounced in arithmetic and Clifford-dominated families, such as the CDKM ripple-carry adder, full adder, modular adder, BV, and GHZ circuits. For these families, the average improvement at the family-level remains in the upper 30\% to upper 60\% range, while individual instances often achieve improvements between 80\% and 98\%. In such circuits, the total cost is governed by a single pathway, which causes the uniform baseline to allocate resources inefficiently to less active mechanisms. As a result, adaptive reallocation produces substantial multiplicative savings. The consistent ranking of these families at the top of improvement spectrum indicates that this effect is a consequence of the circuit structure rather than a property of optimization methods.

\paragraph{Rotation-heavy circuits}
A comparable trend is evident in rotation-heavy circuit families such as QFT, QFT-entangled, QPE-exact, and QPE-inexact. For these families, the average improvements at the family-level range from the upper 20\% to the lower 30\% for GT, with SA and QPSO achieving marginally higher values. Individual instances may attain improvements of 50\% to 65\%, and in favorable cases, up to 70\% to 80\%. In these circuits, rotation synthesis constitutes a substantial fraction of the total cost. Accordingly, reallocating resources to the rotation pathway reduces the required synthesis precision and shortens the decompositions throughout the circuit. This structural feature is present for all three optimization methods, with SA and QPSO recovering a somewhat greater proportion of the available improvement.

\paragraph{Variational and parameterized circuits}
Variational circuit families display the greatest heterogeneity in achievable improvement. QAOA, QNN, VQE Su2, and VQE Two Local exhibit a broad range of instance-wise improvements, from near zero to above 60\%, and in some cases approaching 70\% to 80\%. At the family-level, average improvements are typically in the high 20\% to high 30\% range. This wide dispersion is attributable to structural factors intrinsic to the circuits, rather than to stochastic variation in the optimization process. In particular, the attainable gain is highly sensitive to instance-level parameters, including qubit count, ansatz depth, the range of parameterized gates, and the resulting gate composition. Shallow or weakly parameterized instances may remain dominated by a single mechanism and thus permit substantial savings, whereas deeper or more densely parameterized instances tend to exhibit a more balanced cost profile and consequently yield smaller improvements. By contrast, VQE Real Amplitudes demonstrates a narrower distribution of improvements, reflecting the greater regularity of its gate structure across circuit sizes.

\paragraph{Balanced-profile circuits}
Balanced-profile circuit families, including Shor, Grover, multiplier, and quantum walk, consistently exhibit modest improvements, remaining near the lower end of the observed spectrum. Across all three optimization methods, the gains are limited, with SA and QPSO achieving only marginally greater improvements than GT, and most of the improvement at the instance-level remains below 40\%. In these circuits, the contributions from logical operations, T-state distillation, and rotation synthesis are approximately balanced, resulting in a uniform baseline that is already close to an efficient configuration. Consequently, the limited improvements reflect a lack of structural flexibility within these circuits, rather than any deficiency in the optimization methods themselves. The consistency of these low improvements is significant, as it indicates that the primary determinant of achievable gains is the inherent structure of the allocation problem for each circuit family.
\vspace{-1in}
\paragraph{Random and application-specific circuits}
Random and application-driven circuit families occupy an intermediate region within the improvement spectrum. For the Random Circuit and BMW families, the average improvements at the family-level typically fall within the upper 20\% to upper 30\% range. However, individual instances display a broad range of improvements, from near zero to above 50\%, and in some cases exceeding 60\%. This variability aligns with the presence of moderately imbalanced error profiles that differ substantially across instances. Importantly, these families lack a characteristic level of improvement; instead, the achievable gains are determined by the specific gate composition and the resulting error profile of each circuit instance.

\begin{tcolorbox}[
    colback=gray!10,
    colframe=white,
    left=2mm,
    enhanced,
    borderline west={2pt}{0pt}{teal}
]
\textbf{RQ2.} The magnitude of improvement is governed primarily by asymmetry in the underlying error profile. Circuits dominated by a single error mechanism achieve the largest gains (80-98\%) because uniform allocation wastes budget on inactive pathways, whereas balanced circuits show the smallest gains (0-30\%) because the uniform baseline already lies near the optimum; variational circuits span the full range because the achievable benefit depends on the qubit count, ansatz depth, and resulting gate composition of each circuit instance.
\end{tcolorbox}

\subsection{Influence of Structural Circuit Features}
\label{structural}
\begin{figure*}[!t]
    \centering
    \includegraphics[width=1\linewidth]{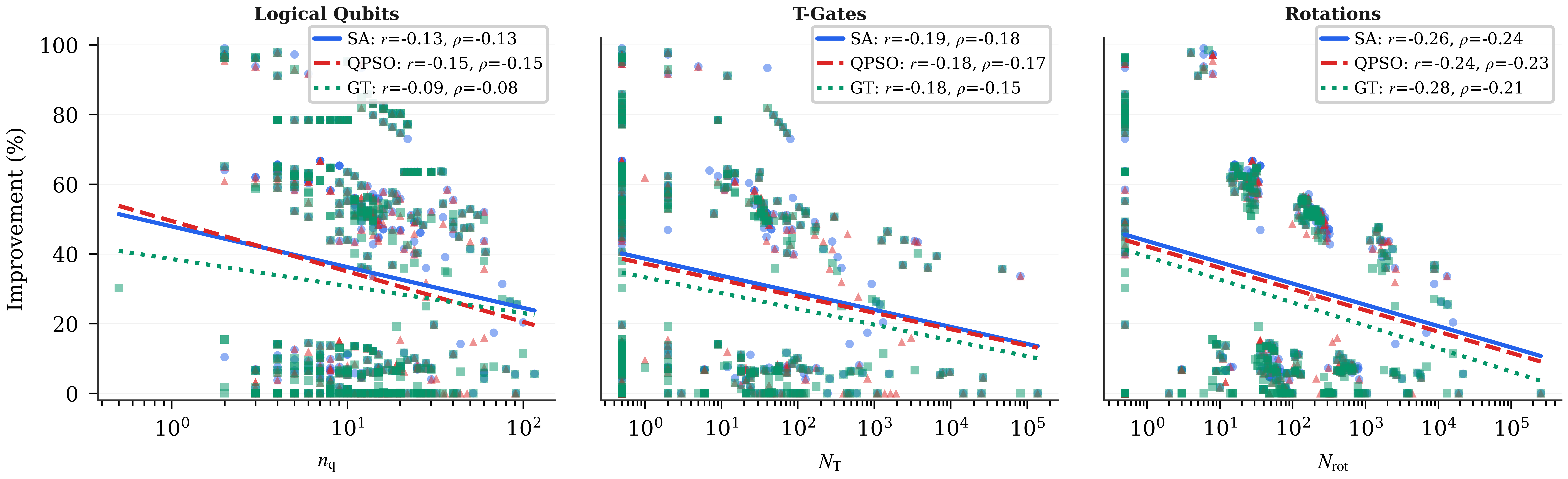}
    % \vspace{-0.3in}
    \caption{Improvement against logical qubit count, T-gate count, and rotation count. Pearson ($r$) and Spearman ($\rho$) correlations are reported in each panel.}
    \label{fig:structural}
\end{figure*}

Figure~\ref{fig:structural} evaluates the extent to which the magnitude of improvement can be anticipated from three pre-optimization circuit descriptors: the number of logical qubits \(n_q\), the T-gate count \(N_T\), and the rotation count \(N_{\mathrm{rot}}\). For each descriptor, improvement is plotted on a logarithmic scale for SA, QPSO, and GT, with both Pearson correlation coefficient \(r\) and the Spearman rank correlation \(\rho\) reported. The results indicate that the features of the structural circuit are only weakly negatively correlated with the improvement in all methods. Specifically, for logical qubit count, the correlations range from \((r,\rho)=(-0.09,-0.08)\) for GT to \((-0.15,-0.15)\) for QPSO. For T-gate count, the correlations are similarly small, ranging from \((-0.18,-0.15)\) to \((-0.19,-0.18)\). Rotation count exhibits the strongest association among the three descriptors, but the effect remains limited, with correlations spanning approximately \((-0.24,-0.23)\) to \((-0.28,-0.21)\). These findings suggest that while larger or more gate-intensive circuits may experience marginally reduced improvement from budget reallocation, the overall association is consistent across SA, QPSO, and GT.

The observed weak dependence is itself a significant finding. If simple structural descriptors are strong predictors of optimization gain, one would expect to observe pronounced monotonic trends and a tighter clustering of data points around the regression lines in each panel. However, the data for all three methods exhibit substantial dispersion. Both small and large circuits are represented across the full range of improvement, as are circuits with moderate and high T-gate or rotation counts. The marginally stronger negative correlation for rotation count indicates that rotation synthesis exerts a somewhat greater influence on the optimization landscape compared to qubit count or T-gate count, but this effect remains insufficient to serve as a robust screening criterion.

Across all three panels, the most prominent feature is the pronounced bimodal distribution of improvement values, rather than any regression trend. For each of the descriptors, namely logical qubit count, T-gate count, and rotation count, the data are separated into two distinct bands: a high-improvement band spanning approximately 40\% to 99\%, and a low-improvement band from 0\% to 15\%, with few instances in the intermediate range. This bimodality is consistent across SA, QPSO, and GT. Both small circuits (fewer than ten logical qubits) and circuits with $10^4$ or more rotations appear in both bands. This separation reflects the discrete nature of the resource-estimation process: a reallocation either crosses a threshold that reduces code distance or synthesis overhead, leading to a substantial reduction in resource cost, or it does not, resulting in only marginal improvement. Intermediate cases are rare because code distance is integer-valued, and changes to it induce stepwise shifts in resource requirements.

As a result, no simple screening criterion based solely on circuit dimensions can reliably identify which instances will achieve substantial improvement. For instance, a circuit with $10^5$ rotations may belong to either the high or low improvement band, depending on the interplay among the three error mechanisms that contribute to the total resource cost. This variability is not captured by any single scalar measure of circuit size. The data across all three panels indicate that structural circuit features alone provide limited predictive power for optimization gain. The consistency of this result across SA, QPSO, and GT supports the conclusion that the primary factor governing improvement is the interaction between the circuit's error profile and the estimator, rather than the circuit's absolute scale. This observation motivates the development of predictors that directly quantify allocation imbalance, rather than relying on pre-optimization circuit counts.

\subsection{Budget Allocation Inequality as a Predictor}

The preceding analysis establishes that improvement is governed by the asymmetry of the error profile rather than by circuit scale, but does not provide a single quantitative measure of this asymmetry. To this end, we compute the Gini coefficient of the optimized allocation $\mathbf{s}^*$, a standard measure of distributional inequality that equals zero for the uniform allocation and approaches $(d-1)/d$ when the budget is
concentrated on a single mechanism.

% \begin{figure}[!t]
%     \centering
%     \includegraphics[width=1\linewidth]{fig3_gini_SA_vs_QPSO_vs_GT.png}
%     \vspace{-0.3in}
%     \caption{Improvement against the Gini coefficient of the optimized allocation.}
%     \label{fig:gini}
% \end{figure}

\begin{figure}[!t]
    \centering
    \includegraphics[width=1\linewidth]{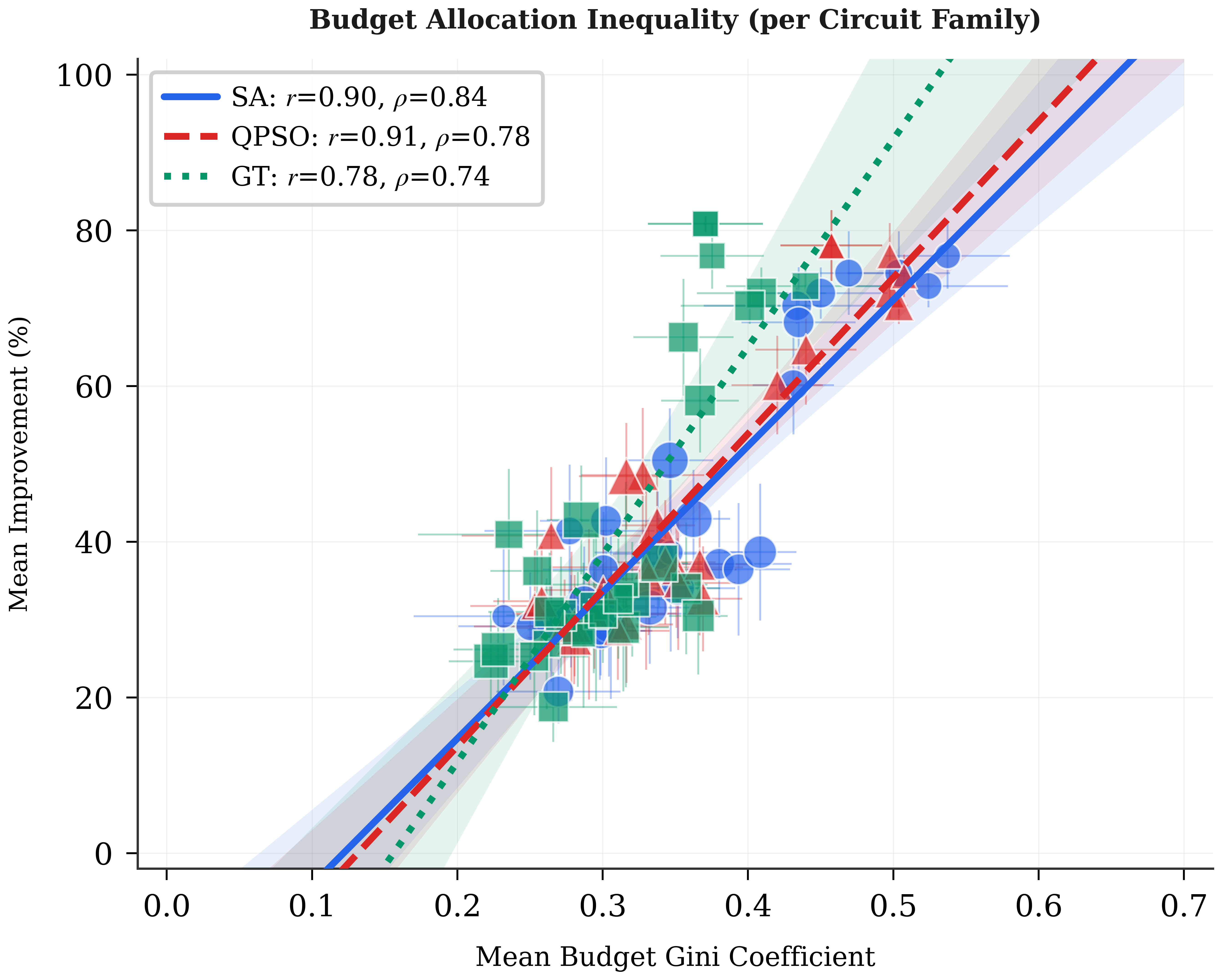}
    % \vspace{-0.3in}
    \caption{Improvement against the Gini coefficient of the optimized allocation.}
     % \vspace{-0.15in}
    \label{fig:gini}
\end{figure}

Figure~\ref{fig:gini} presents the relationship between the Gini coefficient of the optimized error budget allocation and the resulting metric improvement. Circuits within the same MQT benchmark family share a common gate topology and differ only in qubit count. Consequently, their transpiled gate structures, error budget compositions, and optimal allocations are inherently correlated. Treating these correlated instances as independent observations would artificially inflate the effective sample size and exaggerate the statistical significance. Both quantities are therefore aggregated to family-level means ($k=31$ families), ensuring that each data point represents a structurally distinct circuit architecture and that the reported correlations reflect genuine between family variation. The structural analysis in Subsection~\ref{structural} retains instance-level reporting because the primary interest there is the scaling of improvement with circuit size within each family, a within family effect that aggregation would obscure. The Gini analysis targets a between family effect, namely whether architecturally distinct circuit families with more asymmetric error budgets achieve systematically greater improvement, and family-level aggregation is the appropriate unit of analysis for this comparison.

The resulting correlations are strong and consistent across all three methods. SA yields $r=0.90$ and $\rho=0.84$, while QPSO yields $r=0.91$ and $\rho=0.78$, and GT yields $r=0.78$ and $\rho=0.74$, all with $p \ll 0.001$.  These values substantially exceed every structural predictor examined in Subsection~\ref{structural} and hold regardless of the underlying optimization strategy. The regression lines for SA and QPSO nearly overlap across the entire Gini range, with closely aligned slopes and broadly overlapping confidence intervals, while GT follows the same positive trend with a somewhat shallower slope. This indicates that the relationship between budget inequality and improvement is primarily determined by circuit architecture, while SA and QPSO recover somewhat larger gains from the same underlying asymmetry.

The attainable improvement increases systematically over the observed Gini range of approximately $0.2$ to $0.55$, rising from $20$--$30\%$ for families such as Grover, Shor, and the integer multiplier to $70$--$80\%$ for highly asymmetric families such as the CDKM ripple carry adder, full adder, and modular adder. This progression is visible for SA, QPSO, and GT alike, although the SA and QPSO trends lie consistently above GT across most of the observed range. The trend is approximately linear throughout, with no evidence of saturation. This progression has a direct interpretation. When the optimal allocation is relatively balanced, the uniform baseline already lies close to the cost minimum on the allocation simplex $\mathcal{S}_\varepsilon$, leaving limited room for reduction in $\mathcal{J}(\mathbf{s})$. When the optimal allocation is strongly unequal, the uniform baseline assigns substantial budget to inactive pathways, and the optimizer eliminates this misallocation by redirecting budget toward the active mechanisms, triggering code distance reductions that produce multiplicative savings in the space-time cost.

The Gini coefficient is computed from the optimized allocation $\mathbf{s}^*$ and therefore serves as a post hoc diagnostic rather than a predictive feature. It cannot be evaluated before optimization and does not replace the optimizer. Its value lies in providing a compact and interpretable explanation for the large cross family variation in improvement observed under SA, QPSO and GT, and in confirming that the gains reported in Table~\ref{tab:comparison} are systematic consequences of error profile asymmetry rather than artifacts of any particular search procedure.

\begin{tcolorbox}[
    colback=gray!10,
    colframe=white,
    left=2mm,
    enhanced,
    borderline west={2pt}{0pt}{teal}
]
\textbf{RQ3.} Structural circuit features alone demonstrate limited predictive value,  whereas the Gini coefficient at the circuit family-level offers a substantially stronger post hoc explanation by quantifying the inequality in allocation associated with optimization gains.
\end{tcolorbox}

% Taken together, the analyses in Figures~\ref{fig:violin}, \ref{fig:structural}, and~\ref{fig:gini} establish three findings. The optimizer achieves a mean space-time cost reduction of 35.33\% across the benchmark suite, with individual families reaching improvements above 90\%. Structural descriptors are poor predictors of this gain, with correlations no stronger than $|r| = 0.28$ and $|\rho| = 0.21$, because improvement is governed by error profile asymmetry rather than circuit scale. The Gini coefficient of the optimized allocation captures this asymmetry directly at
% $r = 0.82$ and $\rho = 0.77$, providing both a diagnostic tool for
% practitioners and evidence that adaptive error budget allocation should be a standard step in any fault tolerant resource estimation pipeline.

\subsection{Comparison with Prior Work and Optimizer Robustness}

Table~\ref{tab:comparison} compares the present results with the two most relevant prior approaches to error budget optimization for fault-tolerant resource estimation, namely the supervised method of Forster et al.~\cite{forster2025improving} and the GT method of Asif et al.~\cite{ronggon2026gametheoreticapproachoptimizing}. The comparison with the GT method is direct, as both approaches are evaluated on the same MQT Bench suite using the same uniform allocation baseline. In contrast, the comparison with the supervised method is not a circuit-by-circuit match, since the benchmark suite has expanded since that study and now includes additional circuit families. However, this comparison remains a meaningful point of reference because both studies address the same error budget allocation task within the same benchmark framework and report improvements relative to the same default baseline.

\begin{table}[t]
\centering
\caption{Resource improvement over uniform budget allocation. Mean and standard deviation are reported across four independent random seeds. Wall-clock time is for the full 433-circuit suite on Kaggle.}
\label{tab:comparison}
\begin{tabular}{C{2.2cm} C{0.5cm} C{1.5cm} C{0.8cm} C{1.2cm}}
\hline
\textbf{Methodology} & \textbf{Circuits} & \textbf{Average} & \textbf{Maximum} & \textbf{Time} \\
\hline
Forster et al.~\cite{forster2025improving} & 383 & 15.6\% & 77.7\% & --- \\
Asif et al.~\cite{ronggon2026gametheoreticapproachoptimizing} & 433 & 30.22\% & 97.81\% & $\sim$287 min  \\
This work (SA)   & 433 & 35.18$\pm$0.11\% & 97.81\% & $\sim$20 min \\
This work (QPSO) & 433 & 33.84$\pm$0.08\% & 97.81\% & $\sim$25 min \\
\hline
\end{tabular}
\end{table}

Within this comparative framework, Table~\ref{tab:comparison} demonstrates both an increase in average improvement and a significant enhancement in computational efficiency. The supervised method achieves an average improvement of 15.6\%, whereas the GT method increases this figure to 30.22\% on the 433-circuit suite. The present SA and QPSO frameworks further advance these results, attaining average improvements of $35.18 \pm 0.11\%$ and $33.84 \pm 0.08\%$, respectively. Notably, these gains are accompanied by a reduction in total runtime for the full benchmark suite from approximately 287 minutes with the GT method to approximately 20-25 minutes. These findings indicate that the observed resource savings are achieved without incurring additional optimization overhead.

Timing results play a pivotal role in software workflows for fault-tolerant resource estimation, as the practical value of allocation quality depends on the timely delivery of solutions within iterative design and compilation flows. Our empirical evaluation demonstrates that the proposed framework advances the trade-off between solution quality and computational efficiency, yielding greater average improvements than the same-suite GT approach while reducing runtime by more than an order of magnitude. In contrast to supervised methods, our framework obviates the need for offline dataset construction, model training, or prior circuit exposure, thereby streamlining the optimization process and increasing its suitability for estimator-driven workflows.

The close correspondence between SA and QPSO further substantiates this interpretation. Their mean improvements differ by only 1.34 percentage points, and both achieve the same maximum reduction of 97.81\%, completing the full benchmark suite within a narrow runtime interval. This consistency indicates that the observed gains are attributable to the underlying structure of the allocation problem rather than reliance on a particular optimizer, and supports the use of adaptive error-budget allocation as a practical optimization layer within fault-tolerant resource estimation workflows.

\begin{tcolorbox}[
    colback=gray!10,
    colframe=white,
    left=2mm,
    enhanced,
    borderline west={2pt}{0pt}{teal}
]
\textbf{RQ1.} Implementing error budget allocation as an instance-specific optimization step within the fault-tolerant resource estimation workflow, instead of using a fixed uniform default, consistently reduces estimated space-time overhead by over 33\% on average across diverse benchmark circuits. This enhancement does not require offline training data or gradient information.
\end{tcolorbox}

\begin{tcolorbox}[
    colback=gray!10,
    colframe=white,
    left=2mm,
    enhanced,
    borderline west={2pt}{0pt}{teal}
]
\textbf{RQ4.} The close agreement between the two evaluated metaheuristic optimizers, with average improvements varying by only 1.34 percentage points, indicates that the observed improvement results from the structure of the allocation problem rather than from any single search strategy.
\end{tcolorbox}

\section{Conclusion}
\label{conclusion}

In this work, we present a system-software optimization framework that reduces the average estimated space-time overhead by more than 33\% across 433 circuits from 31 families through adaptive, instance-specific error budget allocation. Compared to the fixed uniform baseline, our approach integrates an optimization layer around the resource estimator and demonstrates consistent improvements across the two tested metaheuristic optimizers, as evidenced by agreement within 1.34 percentage points. These results indicate that substantial quantum resource savings can be achieved without relying on offline training data or prior circuit exposure.

Further analysis indicates that the primary advantage of adaptive allocation arises from asymmetry in the underlying error profile, whereas structural features such as logical qubit count, T-count, and rotation count provide limited explanatory power. This finding explains why fixed uniform allocation often performs suboptimally across diverse circuit families. Consequently, adaptive error budget allocation functions as a practical optimization layer for fault-tolerant resource estimation workflows, enhancing instance-level allocation decisions without altering estimator internals. Future research will focus on integrating adaptive error budget allocation more tightly into fault-tolerant compilation flows and expanding the current formulation to a broader system-aware context that incorporates routing, scheduling, and memory overhead.

\section{Acknowledgments}
This effort is sponsored in part by LA Board of Regents (BOR) and NSF under grant numbers AWD-AM260479 and 2608182, respectively. The U.S. Government is authorized to reproduce and distribute reprints for Governmental purposes, notwithstanding any copyright notation thereon. The views
and conclusions contained herein are those of the authors and should not be interpreted as necessarily representing the official policies or
endorsements, either expressed or implied, of the U.S. Government.

\bibliographystyle{IEEEtran}
\bibliography{ref}

% Place \IEEEpubid right before the column adjustment and bibliography
\IEEEpubid{Accepted for publication at IEEE Quantum Week 2026.}
\IEEEpubidadjcol

\end{document}